\documentclass[prl,aps,twocolumn]{revtex4}

\usepackage{times}
\usepackage{graphicx}
\usepackage{float}
\usepackage{latexsym,amsmath,amssymb,bm,euscript}
\usepackage{color}
\usepackage{subfigure}
\usepackage{epstopdf}
\usepackage[colorlinks=true,linkcolor=blue,citecolor=blue]{hyperref}
\usepackage{type1cm}

\usepackage{appendix}

\begin{document}
\title{The Fractional Quantum Hall States  at $\nu=13/5$ and $12/5$ and their Non-Abelian  Nature}
\author{W. Zhu$^{1,2}$, S. S. Gong$^2$, F. D. M. Haldane$^1$ and D. N. Sheng$^2$}
\affiliation{$^1$Department of Physics, Princeton University, Princeton, NJ 08544, USA}
\affiliation{$^2$Department of Physics and Astronomy, California State University, Northridge, CA 91330, USA}

\begin{abstract}
Topological quantum states with non-Abelian Fibonacci anyonic excitations
are widely sought after for the exotic fundamental physics they would
exhibit,  and for universal quantum computing applications.
The fractional quantum Hall (FQH) state at filling factor $\nu=12/5$ is  a promising candidate,
however, its precise nature is still under debate and no consensus has been achieved so far.
Here, we investigate the nature of the FQH $\nu=13/5$ state and its particle-hole conjugate state at $12/5$
with the  Coulomb interaction, and address the issue of possible competing states.
Based on a large-scale density-matrix renormalization group (DMRG) calculation in spherical geometry,
we present evidence that the essential  physics of the Coulomb ground state (GS)
at $\nu=13/5$ and $12/5$ is captured by the $k=3$ parafermion  Read-Rezayi state ($\text{RR}_3$),
including  a robust excitation gap and the topological fingerprint from  entanglement spectrum and topological entanglement entropy.
Furthermore, by considering the infinite-cylinder geometry (topologically equivalent to torus geometry),
we expose the non-Abelian GS sector corresponding to a Fibonacci anyonic quasiparticle,
which serves as a signature  of  the $\text{RR}_3$ state at $13/5$ and $12/5$ filling numbers.
\end{abstract}


\maketitle

\textit{Introduction.---}
While fundamental particles in nature are either bosons or fermions,
the emergent excitations in two-dimensional strongly-correlated systems  may
 obey fractional or anyonic statistics \cite{Tsui1982,Laughlin1983}.
After two decades of study \cite{Willett1987,Pan1999,Xia2004,Choi2008,Pan2008,Kumar2010,Radu2008,Dolev2012,Willett2013,Baer2014,Chi2012},
current interest in exotic excitations 
focuses on states of matter with non-Abelian quasiparticle excitations \cite{Moore,Greiter,Read1999},
and their potential applications to the rapidly evolving
field of quantum computation and cryptography \cite{Nayak,Kitaev2003,Freedman2002,Sarma2005,Hormozi2007,Bonesteel2005}.
So far the most promising platform for realization of  non-Abelian statistics
is the fractional quantum Hall (FQH) effect in the first excited  Landau level,
and  two of the most interesting examples are at filling factors $\nu=5/2$ and $12/5$. 
The $\nu=5/2$ state is widely considered to be the candidate for
the Moore-Read state hosting non-Abelian Majorana quasiparticles \cite{Moore,Greiter,Read1999}.
Experiments have revealed that the $12/5$ state  appears to behave
differently from the conventional FQH effect \cite{Xia2004,Kumar2010},
and  may also be a candidate state for hosting non-Abelian excitations.
However, 
the exact nature of the FQH $12/5$ state is still undetermined due to
the existence of other possible competing candidate states.

Several ground-state (GS) wavefunctions have been proposed
\cite{Read1999,Rezayi2009,Bonderson2008,Bonderson2012,Sreejith2011,Sreejith2013}
as models for the observed  FQH effect at $\nu=12/5$ \cite{Xia2004,Kumar2010,Chi2012}.
The most exciting candidate  is
the $k=3$ parafermion state proposed by Read and Rezayi ($\text{RR}_3$)\cite{Read1999}.
This $\text{RR}_3$ state describes
a condensate of three-electron clusters that forms
an incompressible state at $\nu=13/5$ \cite{Read1999}.
One can also construct the particle-hole partner of the RR$_3$ state to describe the $12/5$ FQH effect.
Besides the $\text{RR}_3$ state, some competing candidates for $\nu=13/5$ or
$12/5$ exist: a hierarchy state \cite{Haldane1983,Halperin1984}, a Jain composite-fermion (CF) state \cite{JainBook},
a generalization of the non-Abelian Pfaffian state by Bonderson and Slingerland (BS) \cite{Bonderson2008,Bonderson2012},
and a bipartite CF state \cite{Sreejith2011,Sreejith2013}.
So far, the true nature of the $12/5$ and $13/5$ FQH states  remains undetermined.
The main challenges in settling  this issue are the
limited computational  ability and the lack of an efficient diagnostic method.
For example, from exact diagonalization (ED) calculations in the
limited feasible range of system sizes,
it is found that the overlaps between the Coulomb GS at $\nu=12/5$ and different model wavefunctions are all relatively large \cite{Read1999,Bonderson2012},
while the extrapolated GS energies of the RR$_3$ and BS states are very close in the
thermodynamic limit \cite{Wojs2009,Bonderson2012}. 
Taken as a whole, previous studies have left the nature of the Coulomb GS at $\nu=13/5$ and $12/5$ unsettled.

Recently, there has been growing interest in connecting quantum entanglement \cite{Kitaev2006,Levin2006,Haldane2008,YZhang2012}
with  emergent topological order \cite{Wen1990,Wen1995} in strongly interacting systems,
which  offers a new route to identification of  the precise topological order of a many-body state. 
Although characterization of entanglement has been successfully used
to identify various well-known types of topological order \cite{Haque2007,Lauchli2010,Papic2011,Cincio2013,Zatel2013,Tu2013,HCJiang,WZhu2014},
application of  the method to a system with competing phases still
faces challenges when
ED studies  suffer from strong finite size effects, and other methods such as quantum Monte-Carlo
suffer from sign problems.  The recent development of the high efficiency  density-matrix renormalization group (DMRG)
in momentum space \cite{JizeZhao2011,Zatel2013}
allows the study of such systems in sphere and cylinder geometries,
both of which  can be used to make concrete predictions of the physics of real systems in the thermodynamic limit.
Here  we combine these advances,  and use  these two geometries to
address the long-standing issues of the FQH at $\nu=12/5$ and $13/5$.

In this paper, we study  the FQH at $\nu=12/5$ and $13/5$ filling   by using the
state-of-the-art density-matrix renormalization group (DMRG) numerical simulations.
By studying  large systems up to $N_e=36$ on spherical geometry, we
establish that the Coulomb GS at $\nu=13/5$ is an incompressible FQH state,
protected by a robust neutral excitation gap $\Delta_n\approx 0.012 (e^2/l_B)$.
Crucially,
we  show that the entanglement spectrum (ES)  fits
the corresponding $SU(2)_3$ conformal field theory (CFT)
which describes the edge structure of the parafermion $\text{RR}_3$  state.
The topological entanglement entropy (TEE) is also
consistent with the predicted value for the $\text{RR}_3$ state, indicating the emergence of
Fibonacci anyonic quasiparticles.
Moreover, we also perform a finite-size scaling analysis of the GS energies for $\nu=12/5$ states at different
shifts corresponding to the particle-hole-conjugate of the $\text{RR}_3$ state,
the Jain state and BS state.
Finite-size scaling confirms that the ground state with topological shift $\mathcal{S}= -2 (3)$
(where RR$_3$ state is expected to occur) is energetically favored in the thermodynamic limit.
Finally, to explicitly demonstrate the topological degeneracy,
we obtain  two topological distinct  GS sectors  on the  infinite cylinder using infinite-size DMRG. While one
sector is the  identity sector matching to the GS from the sphere,
the new sector  is identified as the non-Abelian sector with a Fibonacci anyonic quasiparticle through its characteristic ES and TEE.
Thus we  establish that the essence of the FQH state at $\nu=13/5$
is fully  captured by the non-Abelian parafermion $\text{RR}_3$
state (and by its particle-hole conjugate at $\nu$ = 12/5)
and show that it is stable against perturbations as we change the
Haldane pseudopotentials and the layer width of the system.

\textit{Model and Method.---}
We use the Haldane representation \cite{Haldane1983,Fano1986,Greiter2011} in which
the $N_e$ electrons are confined on the surface of a sphere
surrounding  a magnetic monopole of strength $Q$.
In this case, the  orbitals of the $n$-th LL 
are  represented as orbitals with azimuthal angular momentum  $-L,-L+1,...,L$,
with $L=Q+n$ being the total angular momentum.
The total magnetic flux through the spherical surface
is  quantized to be an integer $N_s=2L$.
Assuming that electron spins are fully-polarized and neglecting Landau-level mixing,
the Hamiltonian in the spherical geometry can be written as:
\begin{equation*}
H=\frac{1}{2}\sum_{m_1+m_2=m_3+m_4}\langle m_1,m_2|V|m_3,m_4 \rangle \hat{a}^{\dagger}_{m_1}\hat{a}^{\dagger}_{m_2}\hat{a}_{m_3}\hat{a}_{m_4}
\end{equation*}
where $\hat{a}^{\dagger}_m$ ($\hat{a}_m$ )
is the creation (annihilation) operator at the orbital $m$
and $V$ is the Coulomb interaction between electrons in units of $e^2/l_B$ with $l_B$ being the magnetic length.
The two-body Coulomb interaction element can be decomposed as
\begin{equation*}
\langle i,j|V|p,q\rangle=\sum_{l=0}^{2L}\sum_{m=-l}^{l} \langle L,i;L,j|l,m\rangle \langle l,m |L,p;L,q\rangle \mathcal{V}^n(l)
\end{equation*}
where $\langle L,i;L,j|l,m\rangle$ is the Clebsch-Gordan coefficients
and $\mathcal{V}^n(l)$ is the  Haldane pseudopotential
representing the pair energy of two electrons with relative angular momentum $2L-l$ in $n$-th LL \cite{Haldane1983,supple}.
For electrons at fractional filling factor $\nu$,
$N_s=\nu^{-1}N_e-\mathcal{S}$, where $\mathcal{S}$ is the
curvature-induced ``shift'' on the sphere.

Our calculation is based on the unbiased DMRG method \cite{White,Xiang1996,Shibata,Feiguin2008,JizeZhao2011,Hu2012},
combined with ED. The (angular) momentum-space DMRG allows us to use the total electron number $N_e$ and
the total z-component of angular momentum $L^{tot}_z=\sum_{i=1}^{N_e}m_i$ as
good quantum numbers to reduce the Hilbert subspace dimension \cite{JizeZhao2011}.
Here, we report the result at $\nu=13/5 (12/5)$ with
electron number up to $N_e=36 (22)$ by keeping up to
$30000$ states with optimized  DMRG, which allows
us to obtain accurate results for energy and the ES
on much larger system sizes beyond the ED limit ($N^{ED}_e=24(16)$ at $\nu=13/5 (12/5)$).

\begin{figure}[t]
\includegraphics[width=0.5\textwidth]{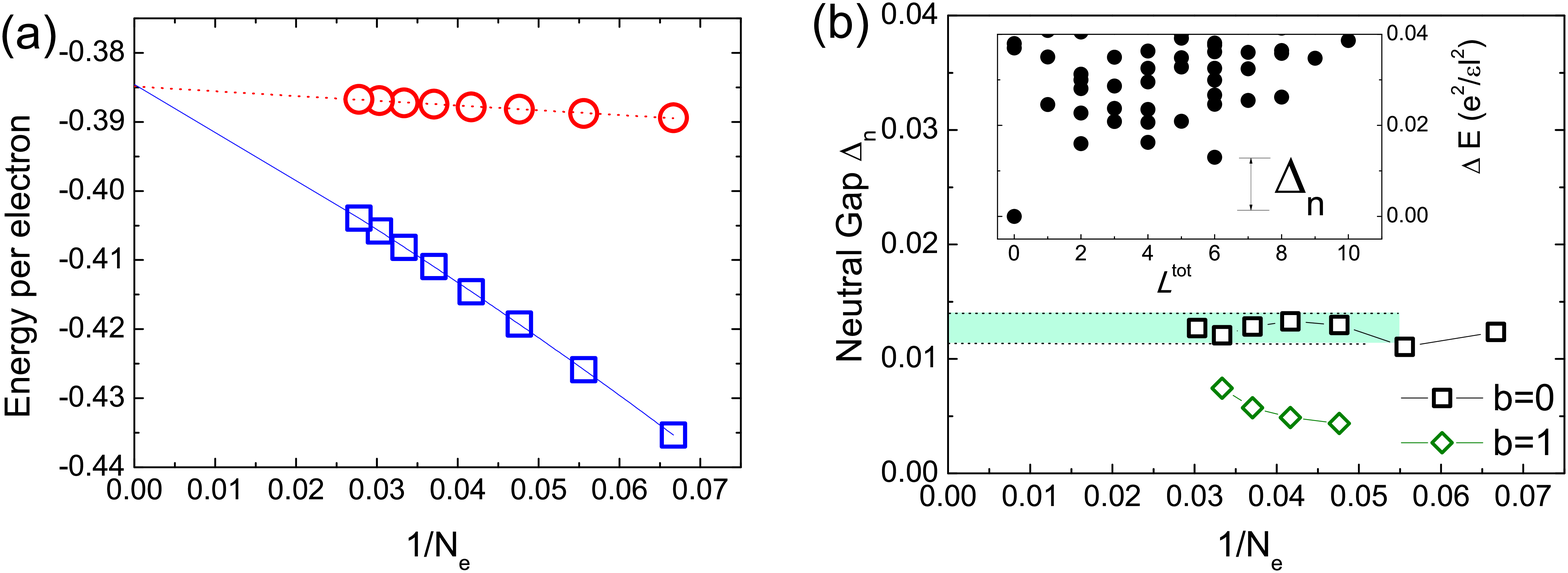}
\caption{(a) The groundstate energy per electron (blue dots) corresponding to the $\nu = 13/5$ state. The blue line shows the extrapolated values obtained using a quadratic function of $1/N_e$. The red dots shows the rescaling energy by a renormalized magnetic length
and the red line is the linear fitting. (b) The neutral gap $\Delta_n$ for
$13/5$ state as a function of the  $1/N_e$ ($b$ is the layer-width parameter \cite{supple}).
Inset: Energy spectrum versus total angular momentum $L^{tot}$ for $N_e=21$. $\Delta_n$ is
defined as the energy difference between the lowest energy state (in $L^{tot}=0$) and the
first excited state (in $L^{tot}\neq0$).} \label{energy}
\end{figure}

\textit{Groundstate Energy, Energy Spectrum  and Neutral Gap.---}
We first compute the GS
energies for a number of systems  up to $N_e=36$ at $\nu=13/5$,
with a shift $\mathcal{S}=3$ consistent with the $\text{RR}_3$ state.
As shown in the low-lying energy spectrum in the inset of Fig. \ref{energy}(b) obtained from ED for $N_e=21$,
the GS is located in the $L^{tot}=0$ sector and  is separated from
the higher energy continuum by a finite gap, which signals an
incompressible FQH state. The extrapolation of the GS
energy to the thermodynamic limit can be carried out using a
quadratic function of $1/N_e$ (blue line), or a linear fit in $1/N_e$
(red line) after renormalizing the energy by $\sqrt{2Q\nu/N_e}$ to take
into account the curvature of the sphere \cite{Morf1985}, as shown in Fig.
1(a). We obtain the $E_0/N_e= -0.38458(24)$ (blue line) and
$-0.38487(9)$ (red line), which demonstrates consistency between
the two extrapolating schemes.

We also calculated the neutral excitation gap $\Delta_n$ at $\nu = 13/5$ \cite{note1}. 
This is equivalent to the energy difference between
the GS and the ``roton minimum''\cite{Haldane1985,GMP,BoYang2012} as illustrated
in the inset of Fig. \ref{energy}(b). The roton minimum corresponds to
the lowest
excitation energy of a  quasielectron-quasihole pair \cite{BoYang2012}.
Fig. \ref{energy}(b) shows $\Delta_n$ as a function of $1/N_e$, where
the large-system results
indicate that the neutral gap approaches a nonzero value $\Delta_n\approx 0.012\pm0.001$ for $N_e\geq21$.
Since the hamiltonian in this paper is particle-hole symmetric, the neutral gap at $\nu=12/5$ and $13/5$ are expected to be identical \cite{note2}.
In addition, if the effect of finite layer-width is considered\cite{supple}, the neutral-excitation gap is reduced but still remains consistent with a nonzero value (Fig. \ref{energy}(b)).

\begin{figure}[!htb]
\includegraphics[width=0.35\textwidth]{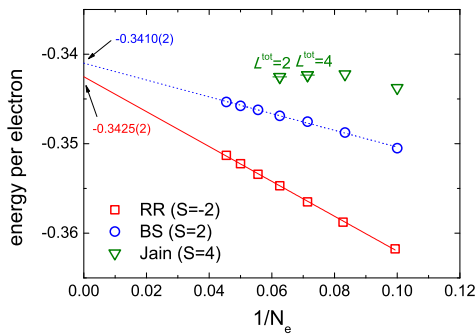}
\caption{Finite-size extrapolation of the ground-state (GS) energies for different shifts corresponding to  different candidate states  at $\nu=12/5$.
All energies have been rescaled by the renormalized magnetic length.
The angular momentum of the GS is shown whenever it is nonzero ($L^{tot}\neq0$).
} \label{vs}
\end{figure}

\textit{Competing states.---}
In Fig. \ref{vs}, we compare the GS
energies per electron of three known candidates for $\nu = 12/5$:
the particle-hole conjugate of the  $\text{RR}_3$  state with a shift $\mathcal{S}=-2$, the non-Abelian BS state with $\mathcal{S}=2$
\cite{Bonderson2008}, and Jain state with $\mathcal{S}=4$.
We  find that the lowest-energy state for the Jain state shift ($\mathcal{S}=4$) in larger system sizes
has a  total angular momentum  $L^{tot}\neq0$,
indicating that it represents excitations of some other incompressible state rather than
the Coulomb GS at $\nu=12/5$ \cite{Sreejith2013}.
Secondly, the GSs with the $\text{RR}_3$ and BS shifts continue to have $L^{tot}=0$ for the systems that we have studied,
and the extrapolation based on the result for $10\leq N_e\leq 22$
leads to   $E_0/N_e=-0.3425$ for the $\text{RR}_3$ state and
$E_0/N_e=-0.3410$ for the BS state, respectively.
Compared to the previous studies \cite{Bonderson2012,Wojs2009},
the extrapolation errors are reduced  by the inclusion of
larger system sizes obtained using DMRG.
Our calculations suggest that
the $\text{RR}_3$ state with  shift $\mathcal{S}=-2$($\mathcal{S}=3$) is energetically favored as the GS at $\nu=12/5(13/5)$.
Our results are consistent with the interpretation that
the $\text{RR}_3$ state describes the true GS (see the full evidence below),
while the other states at nearby shifts correspond to  states with
quasiparticle or quasihole excitations.

\textit{Orbital ES.---} Li and Haldane first established that the orbital ES
of the GS of FQH phase contains information about
the counting of their edge modes \cite{Haldane2008,Wen1995}.
Thus, the orbital ES provides a ``fingerprint'' of the topological order, which can
be used to identify the emergent topological phase in a microscopic Hamiltonian
\cite{Haldane2008,Lauchli2010,Papic2011,Cincio2013,Zatel2013}.

\begin{figure}[t]
 \begin{minipage}{0.98\linewidth}
 \includegraphics[width=3.0in]{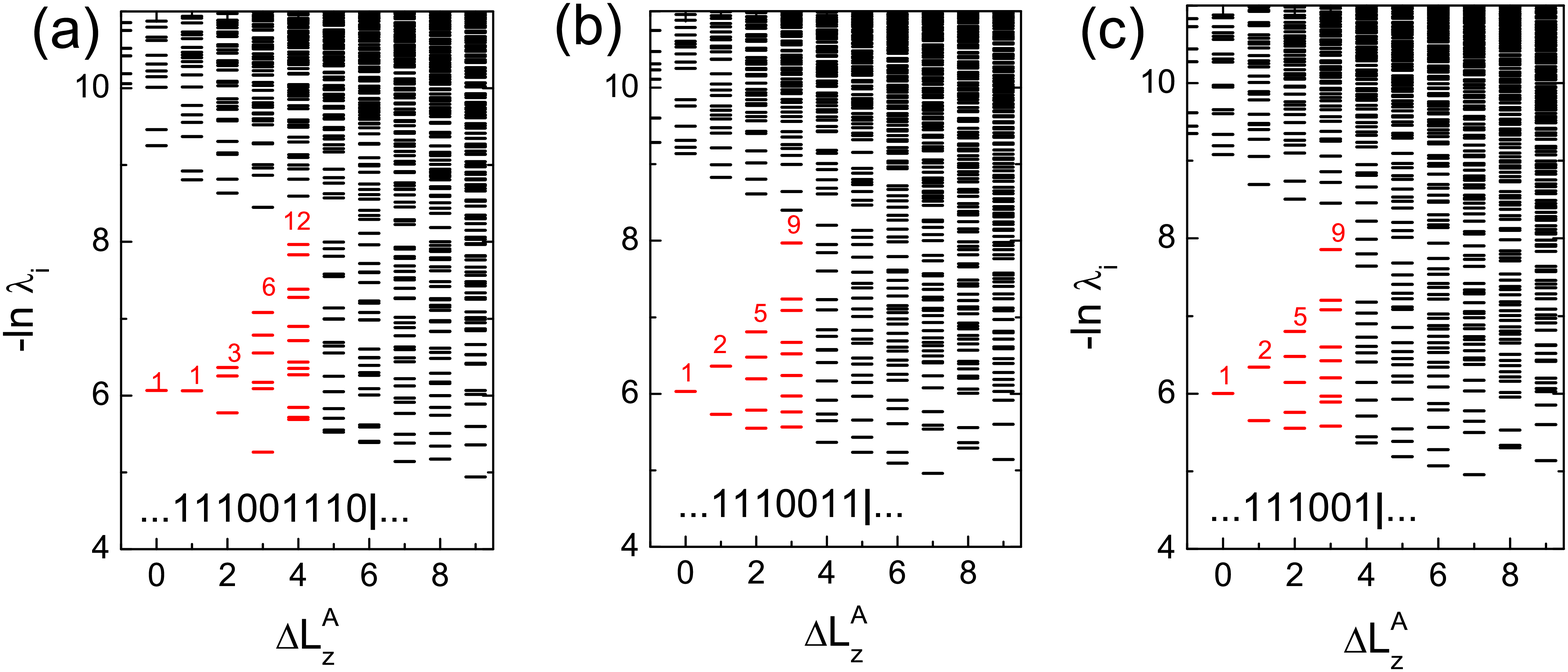}
 \includegraphics[width=2.8in]{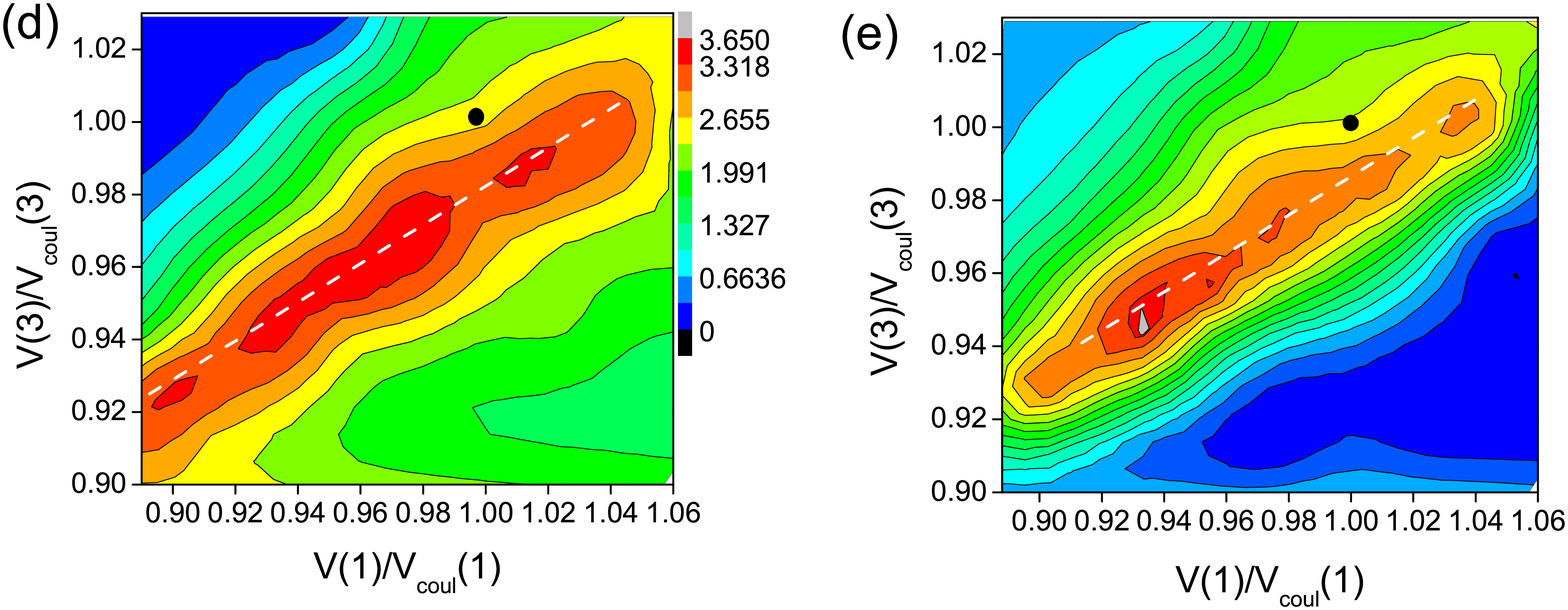}
 \end{minipage}
 \caption{(a-c) The low-lying orbital ES of $N_e=36$ are shown for three
different partitions. The lower ES level counting in the sector
$\Delta L^A_z = 0,1,2,3,4$ are labeled by color, where $\Delta L^A_z=L^A_z-L^A_{z,min}$ with
$L^A_{z,min}$ as the quantum number where the primary field occurs.
The entanglement gap of orbital ES of $N_e=24$ is shown for partition (d) with $3n$ electrons and (e) with $3n+1$ electrons in the subsystem
as a function of pseudopotential $\mathcal{V}^{1}(1)/\mathcal{V}^{1}_{Coul}(1)$ and $\mathcal{V}^{1}(3)/\mathcal{V}^{1}_{Coul}(3)$,
where $\mathcal{V}^{1}_{Coul}(l)$ are the Coulomb values of pseudopotentials.
The black point corresponds to the Coulomb point.
} \label{ES}
\end{figure}

As a model FQH state, the $\text{RR}_3$ parafermion state
can be represented by its  highest-density root configuration
pattern of ``1110011100... 11100111'', corresponding to
a generalized Pauli principle of ``no more than three electrons
in five consecutive orbitals'' \cite{Bergholtz2005,Bernevig2008a,Bernevig2008b}.
Consequently, the orbital ES depends on the number of electrons in the
partitioned subsystem \cite{supple}.
In Fig. \ref{ES}, we show the orbital
ES of three distinct partitions for system size $N_e=36$ for Coulomb GS.
For $3n$ electrons in subsystem (Fig. \ref{ES}(a)), the leading
ES displays the multiplicity-pattern $1,1,3,6,12$
in the first five angular momentum sectors $\Delta L^A_z = 0,1,2,3,4$.
For $3n+1$ or $3n+2$ electrons in subsystem (Fig. \ref{ES}(b-c)),
the ES shows the multiplicity-pattern of $1,2,5,9$ in
the $\Delta L = 0,1,2,3$ momentum sectors. The above characteristic
multiplicity-patterns of the low-lying ES agree with the predicted  edge excitation
spectrum of the $\text{RR}_3$ state obtained either from  its associated
CFT, or the  ``$\le 3$ in 5'' exclusion statistics rule \cite{supple,note3}.

In addition, we vary the Haldane
pseudopotentials $\mathcal{V}^{1}(1)$ and $\mathcal{V}^{1}(3)$
(keeping all others at their Coulomb-interaction values),
and map out  an ES-gap  diagram which illustrates  the robustness of the FQH state
as  the interaction parameters are changed\cite{Morf2010,Peterson2008,Biddle2011,Pakrouski2014}.
In Fig. \ref{ES}, we plot the entanglement gap (for the lowest-$L^z$ ES level)\cite{Haldane2008,JizeZhao2011}
as a function of $\mathcal{V}^{1}(1)/\mathcal{}\mathcal{V}^{1}_{Coul}(1)$ and $\mathcal{V}^{1}(3)/\mathcal{V}^{1}_{Coul}(3)$,
where $\mathcal{V}^{1}_{Coul}(l)$ are the Coulomb values of pseudopotentials.
We find that the entanglement gap is  robust
in a region centered  at an approximately-fixed $\mathcal{V}^{1}(1)/\mathcal{V}^{1}(3)$ ratio (indicated by the white line).
Away from that,  for  the regime $\mathcal{V}^{1}(1)/\mathcal{V}^{1}_{Coul}(1)<0.92$ and $\mathcal{V}^{1}(3)/\mathcal{V}^{1}_{Coul}(3)>0.98$,
we find  a rapid drop of the entanglement gap indicating a quantum phase transition.
We have also studied  the effect of the ES of modifying the Coulomb interaction
with a realistic layer width ($\textit{b}$)  \cite{supple}, and
find that  the $\text{RR}_3$ state persists until $\textit{b}/l_B\sim2$, which is qualitatively consistent
with the results of varying $\mathcal{V}^{1}(1)$ and $\mathcal{V}^{1}(3)$.

\textit{Topological Entanglement Entropy.---}
For a two-dimensional gapped topologically-ordered state, the
dependence of the entanglement entropy $S_A(l_A)$ of the subsystem A
on the finite
boundary-cut length $l_A$ has the form $S_A(l_A)=\alpha l_A - \gamma$,
where TEE $\gamma$ is related to the total quantum dimension
$\mathcal{D}$ by $\gamma = \ln \mathcal{D}$ \cite{Kitaev2006,Levin2006}.
We have extracted the TEE using  our largest system, $N_e=36$ \cite{supple}.
The TEE obtained was $\gamma=1.491\pm0.091$, consistent with
the theoretically-predicted value  $\gamma=\ln \mathcal{D}=\ln
\sqrt{5(1+\phi^2)}\approx1.447$ for the $\text{RR}_3$ state,
where each non-Abelian Fibonacci anyon quasiparticle
contributes an individual quantum dimension $d_F=\phi=(\sqrt{5}+1)/2$ ($\phi$ denotes the Golden Ratio).
The appearance of $d_F=\phi$ is a signal of the emergence  of
Fibonacci anyon quasiparticles, and arises because two Fibonacci
quasiparticles may fuse either into
the  identity or into a single Fibonacci quasiparticle \cite{WZhu2014}.
This exotic property
makes Fibonacci quasiparticles  capable of universal quantum computation \cite{Nayak}.

\begin{figure}[!htb]
 \begin{minipage}{0.98\linewidth}
 \includegraphics[width=3.5in]{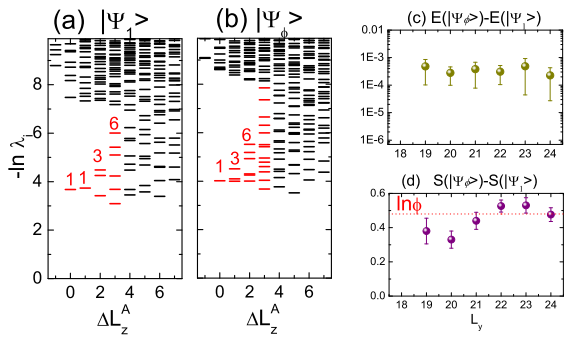}
 \end{minipage}
 \caption{ (a-b) The low-lying orbital ES of $\Psi_1$ and $|\Psi_{\phi}\rangle$ by setting $L_y=24 l_B$.
 $|\Psi_{1(\phi)}\rangle$ denotes the GS with identity $\openone$ (Fibonacci $\phi$) anyonic quasiparticle.
  (c) Energy difference and (d) entropy difference between $|\Psi_{1}\rangle$ and $|\Psi_{\phi}\rangle$,
 obtained from infinite DMRG on cylinder geometry with varying $L_y$.
The error bars are determined based on results from ten different infinite DMRG calculations
for each sector\cite{supple}.
} \label{sector}
\end{figure}

\textit{Topological Degeneracy on the infinite cylinder.---}
Topologically-ordered states have characteristic GS degeneracies
on compactified spaces.
To access the different topological sectors at $\nu=13/5$, we implemented the
infinite-size DMRG in cylinder geometry with a finite circumference  $L_y$ \cite{Zatel2013,Zatel2015,supple}.
For each value of $L_y$,
we repeatedly calculated GSs using different random initializations for the infinite DMRG optimization.
We found that each infinite DMRG simulation converged to one of the two states:
$|\Psi_{1}\rangle$ and $|\Psi_{\phi}\rangle$.
These states are distinguishable by their orbital ES as shown in Fig. \ref{sector}:
$|\Psi_{1}\rangle$ has the same ES structure as in Fig. \ref{ES}(a-c), which
matches the identity sector  with root configuration ``$\ldots 0111001110\ldots$''.
On the other hand,  $|\Psi_{\phi}\rangle$ shows the ES multiplicity pattern $1,3,6,13,...$,
which identifies the spectrum  as that of the Fibonacci non-Abelian
sector with root configuration ``$\ldots 1010110101 \ldots$'' \cite{supple}.
Furthermore, these two groundstates are indeed energetically degenerate,
with an energy-difference per electron  of less than $0.0002$ with $L_y=24l_B$,
while the entropy difference between these two states is around $\Delta S\approx \ln \phi \approx0.48$,
consistent with the quantum dimension of the Fibonacci quasiparticle.
Combining this with the fivefold  center-of-mass degeneracy,
we have obtained all the 10 predicted degenerate $\text{RR}_3$ GSs on
infinite cylinder (or torus).

\textit{Summary and discussion.---}
We have presented what we believe to be compelling evidence that
the essence of the
Coulomb-interaction ground states at $\nu=13/5$ and $12/5$
is indeed captured by the parafermion $k=3$ Read-Rezayi state $\text{RR}_3$,
in which quasiparticles obey non-Abelian ``Fibonacci-anyon'' statistics.
The neutral excitation gap is found to
be a finite value $\Delta_n\approx 0.012 e^2/l_B$ in the thermodynamic limit.
Results for the entanglement spectrum ``fingerprint'' and
the value of the topological entanglement entropy
show that the edge structure and bulk quasiparticle statistics
are consistent with the prediction bases on  the $\text{RR}_3$ state.
Additionally, we find two topologically-degenerate groundstate sectors on the
infinite cylinder,
respectively corresponding to the identity and the Fibonacci anyonic quasiparticle,
which fully confirms the $\text{RR}_3$ state, without input of any
features (such as shift) taken from  the model wavefunction,
that might have biased the calculation.
The current work opens up a number of directions deserving further exploration.
For example, while the FQH $\nu=12/5$ state
has been observed in experiment, there is no evidence
of a FQH phase at $\nu=13/5$ in the same systems \cite{Xia2004,Kumar2010}.
So far it is not clear whether this
absence is due to a broken particle-hole symmetry from   Landau level mixing, or
other asymmetry effects such as  differences in the  quantum wells \cite{Pan2008}.
Our numerical studies suggest that the outlook for the existence of
such a state at  $13/5$ is promising, and
some positive signs of this may have already been observed  very recently \cite{ZnO2015}.
Numerical studies may also further suggest how
various other  exotic FQH states in the second Landau level at
different filling-factors may be stabilized.

\textit{Note added.---}
After the completion of this work, we
became aware of overlapping results in Refs. \cite{Mong2015}.

WZ thanks Z. Liu for fruitful discussion,
N. Regnault and A. W\'{o}js for useful comments.
We also thank X. G. Wen for stimulating discussion and
M. Zaletel, R. S. K. Mong, F. Pollamann for private communication prior to publication.
This work is supported by the U.S. Department of Energy,
Office of Basic Energy Sciences under grants No. DE-FG02-06ER46305
(WZ, DNS) and  DE-SC0002140 (FDMH), and the  National
Science Foundation through the grant  DMR-1408560 (SSG).
FDMH also acknowledges support from the W. M. Keck Foundation.
WZ also acknowledges the support from MRSEC DMR-1420541 and PREM DMR-1205734
for a visit to Princeton where this work was completed.

\clearpage
\begin{widetext}
\appendix
\begin{appendices}

In this supplemental material, we provide more details of the
calculation and results which were not given
 in the main text.
In Sec. I, we briefly summarize the Haldane pseudopotentials in disk  and spherical geometries used for calculation in the main text.
In Sec. II, we give a detailed analysis of edge excitations of the fermionic Read-Rezayi (RR) $k=3$ state, based on the root configurations.
In Sec. III, we introduce the effect of finite layer-width and show the evolution of entanglement spectrum (ES) with the change of the layer width.
In Sec. IV, we extract the topological entanglement entropy (TEE) based on the dependence of entropy on the length of the orbital cut.
In Sec. V, we show the ES of the $\nu=12/5$ state, which is a
particle-hole-conjugate state of the $\nu=13/5$ state.
In Sec. VI, we introduce the numerical details of the infinite-size density-matrix renormalization group (DMRG)
algorithm on cylinder geometry.

\section{I. Pseudopotentials}
Haldane\cite{Haldane1983} first pointed out that, any
 rotationally-invariant two-body interaction
can be completely described by
a set of ``pseudopotential'' $\mathcal{V}(m)$  with $m \geq 0$,
if projected onto a single Landau level.
The pseudopotential describes the energy of  a pair of particles in a state of given
relative angular momentum $m$. This formalism turned out to be useful
not just for describing the details of the interaction, but also for
understanding the microscopic conditions of the  fractional quantum Hall (FQH) states in such systems.
Here we focus on the pseudopotential formalism for two-body Coulomb interaction,
in two specific geometries,  the  disk (or plane) and the sphere, respectively.

\subsection{1. Disk (plain) geometry}
%

In disk geometry, the Haldane pseudopotentials have been obtained many places \cite{Haldane1983}:
\begin{eqnarray}\label{pseudo_disk}
\mathcal{V}^{(n)}_m=<n,m|V(r)|n,m> = \int\limits_{0}^{\infty} dq q [L_n(\frac{q^2}{2})]^2 L_m(q^2) e^{-q^2} V(q)
\end{eqnarray}
Here $|n,m\rangle$ is a two electron state with relative (azimuthal)
angular momentum $m$ in the $n$-th Landau level. $L_m(x)$ is a
Laguerre polynomial, and the two-body interaction $V(r)=\int d\textbf{q} V(q) e^{i\textbf{q}\cdot\textbf{r}}$.

For the ideal Coulomb interaction in two dimensions $V(r)=1/ r$ ( in
unit of $\frac{e^2}{\varepsilon l_0}$), we have, for the first Landau level,
\begin{eqnarray*}
\mathcal{V}^{(n=0)}(m)= \frac{\Gamma(m+1/2)}{2m!}
\end{eqnarray*}
and for the second Landau level:
\begin{eqnarray*}
\mathcal{V}^{(n=1)}(m)= \frac{\Gamma(m+1/2)}{2m!} \frac{(m-3/8)(m-11/8)}{(m-1/2)(m-3/2)}
\end{eqnarray*}

\subsection{2. Sphere geometry}
In spherical geometry, we first define the total angular momentum
$L = Q + n$, where
$Q$ is the strength of the  magnetic monopole  in the center of the sphere and $n$ is Landau level index.
The pseudopotential $\mathcal{V}^{n}(l)$ is defined as the interaction energy of a pair of electrons as a function of their pair angular momentum $l$.
In this expression, the pseudopotential for particles in a single Landau level
is evaluated from the matrix element
\begin{equation*}
 \mathcal{V}^{n}(l)= <n,L;l|V(r_1-r_2)|n,L;l>
\end{equation*}
For the Coulomb potential on the sphere, we define the chord distance between two
points on a sphere as
\begin{equation*}
 V(r_1-r_2)= V(|r_1-r_2|)=\frac{1}{R\sqrt{2-2\cos\theta_{12}}}=\frac{1}{R}\sum_n P_n(\cos\theta_{12}),
\end{equation*}
where $P_n(x)$ is a Legendre polynomials.
We omit the detailed   calculations of integrals here
\cite{CNYang,Wooten}, and just present  the final result:
\begin{equation}
  \mathcal{V}^{n}(l) =\\
  \frac{1}{\sqrt{Q}}\sum_{k=0}^{2L} (-)^{2Q+l}(2L+1)^2
    \left\{\begin{array}{ccc}
    l & L & L \\
    k & L & L
  \end{array}\right\}
    \left(\begin{array}{ccc}
    L & k & L \\
    -Q & 0 & Q
  \end{array}\right)^2
\end{equation}
where $\left(\begin{array}{ccc}
    L & k & L \\
    -Q & 0 & Q
  \end{array}\right)$ is the Wigner 3j coefficient and
  $\left\{\begin{array}{ccc}
    l & L & L \\
    k & L & L
  \end{array}\right\}$ is the Wigner 6j coefficient.

\section{II. Edge mode counting based on ``root states''}
Here we analyze the counting rules of the edge spectrum in the different topological sectors of the fermionic Read-Rezayi (RR) $k=3$ state.
For simplicity, our analysis below is based on the highest density
``root configurations''\cite{Bernevig2008a,Bernevig2008b}, which are
also the
only-surviving configurations  in the ``thin-torus''
limit of the FQH effect\cite{Bergholtz2005}.
These obey  characteristic ``fractional exclusion statistics'' rules
that constrain the number of particles allowed in a certain group of
consecutive orbitals, and the ``admissible configurations'' that
obey these rules are in one-to-one correspondence with the states of
the of zero-energy eigenstates of the model Hamiltonian for which the
RR states are the highest-density zero-energy states.
For the the $k=3$ $\nu = 3/5$ RR state, the rules are \cite{Bernevig2008b}: ``not more than
one particle in any orbital'' and ``not more than three particles in
any five consecutive orbitals''.

We first assume that the lowest edge-mode with $\Delta L=0$ relates to the quantum Hall system with an open right edge.
For example, for root configuration ``$\ldots 11100111|000000\ldots $'' on the cylinder,
``$|$'' separates the cylinder into left and right subsystems.
This is the exclusion-statistics analog of a ``filled Dirac sea, with
the particles moved as far to the left as possible, consistent with
the exclusion rules.  (this is essentially the the defining relation
of a Virasoro-primary state).
The edge mode excitations can be obtained by the rightwards
rearrangements of the particles  at the edge, increasing the momentum
above its minimum value.   The excited configurations must still obey
the exclusion rules, and this simple rule gives the multiplicities (or
``characters'') of the spectrum as a function of ``momentum''
(Virasoro level) relative to the ``primary'' ``Dirac sea'' state.
This method gives a simple exclusion-statistics-based method for
obtaining the multiplicities that agrees with the very different and more opaque
methods of conformal field theory (CFT)  based on construction of ``Verma modules''.

There are two different topological sectors for the parafermion $\text{RR}_3$ state. One
has the root configuration ``$\ldots 0111001110\ldots$''
 and the other has ``$\ldots 1010110101\ldots$''.
All possible edge excitations of ``$\ldots 0111001110\ldots $''  at $\Delta L\leq3$ are listed in
Tables \ref{t1} and \ref{t2}, which  relate to the partitions with $3n$ and $3n-1$ electrons, respectively.
The Table \ref{t3} and \ref{t4} show the results for ``$\ldots 1010110101\ldots$''.

\begin{table*}[!htb]
\caption{ In this table, we analyze the counting rule of
the edge excitations in the  ``$\ldots 11100111001110|00\ldots$''  sector,
which has multiplicities $1,1,3,6,\ldots$ at $\Delta L=0,1,2,3,\ldots$.}
\begin{ruledtabular}\label{t1}
\begin{tabular}{cccccccc}
$\Delta L=0$&$\Delta L=1$&$\Delta L=2$&$\Delta L=3$& $\Delta L=4$&\\
\hline
$11100111001110|0000$&$11100111001101|0000$&$11100111001100|1000$&$11100111001100|0100$&$11100111001100|0010$ &\\
                     &                     &$11100111001011|0000$&$11100111001010|1000$&$11100111001010|0100$ &\\
                     &                     &$11100110101101|0000$&$11100111000111|0000$&$11100110101100|0100$ &\\
                     &                     &                     &$11100110101011|0000$&$11100111001001|1000$ &\\
                                     & &                         &$11100110101100|1000$&$11100111000110|1000$ &\\
                                     & &                         &$11010110101101|0000$&$11100110101010|1000$ &\\
                                     & &                         &                     &$11100110100111|0000$  &\\
                                     & &                         &                     &$11100101101011|0000$  &\\
                                     & &                         &                     &$11010110101011|0000$  &\\
                                     & &                         &                     &$11100110011100|1000$  &\\
                                     & &                         &                     &$11010110101100|1000$  &\\
                                     & &                         &                     &$1101011010110101101|0000$  &\\
\end{tabular}\end{ruledtabular}
\end{table*}

\begin{table*}[!htb]
\caption{ In this table, we analyze the counting rule of
the edge excitations in the $\ldots 111001110011|00\ldots$ sector,
which has multiplicities  $1,2,5,9,\ldots$ at $\Delta L=0,1,2,3,\ldots$.}
\begin{ruledtabular}\label{t2}
\begin{tabular}{cccccccc}
$\Delta L=0$&$\Delta L=1$&$\Delta L=2$&$\Delta L=3$&\\
\hline
$111001110011|0000$&$111001110010|1000$&$111001110010|0100$&$111001110010|0010$&\\
                   &$111001101011|0000$&$111001110001|1000$&$111001110001|0100$&\\
                   &                   &$111001101010|0000$&$111001101010|0100$&\\
                   &                   &$111001100111|0000$&$111001101001|1000$&\\
                   &                   &$110101101011|0000$&$111001100110|1000$&\\
                                     & &                   &$111001011010|1000$&\\
                                     & &                   &$110101101010|1000$&\\
                                     & &                   &$110101100111|0000$&\\
                                     & &                   &$110100110101101011|0000$&\\
\end{tabular}
\end{ruledtabular}
\end{table*}


\begin{table*}[!htb]
\caption{ In this table, we analyze the counting rule of
the edge excitations in the $\ldots101011010110101|\ldots$ sector,
which has multiplicities  $1,3,6,13,\ldots$ at $\Delta L=0,1,2,3,\ldots$.}
\begin{ruledtabular}\label{t3}
\begin{tabular}{cccccccc}
$\Delta L=0$&$\Delta L=1$&$\Delta L=2$&$\Delta L=3$&\\
\hline
$101011010110101|0000$&$101011010110100|1000$&$101011010110100|0100$& $101011010110100|0010$&\\
                 &$101011010110011|0000$&$101011010110010|1000$& $101011010110010|0100$&\\
                 &$101011010101101|0000$&$101011010101100|1000$& $101011010101100|0100$&\\
                   &               &$101011010101011|0000$& $101011010110001|1000$&\\
                                 & &$101011001110011|0000$& $101011010101010|1000$&\\
                   &               &$101010110101101|0000$& $101011001110010|1000$&\\
                                     & &             & $101011010011100|1000$&\\
                                     & &             & $101010110101100|1000$&\\
                                     & &             & $101011010100111|0000$&\\
                                     & &             & $101011001101011|0000$&\\
                                     & &             & $100111001110011|0000$&\\
                                     & &             & $101010110101011|0000$&\\
                                     & &             & $011010110101101|0000$&
\end{tabular}
\end{ruledtabular}
\end{table*}

\begin{table*}[!htb]
\caption{ In this table, we analyze the counting rule of
the edge excitations in the $\ldots1010110101100|00\ldots$ sector,
which has multiplicities $1,2,5,10,\ldots$ at $\Delta L=0,1,2,\ldots$.}
\begin{ruledtabular}\label{t4}
\begin{tabular}{cccccccc}
$\Delta L=0$&$\Delta L=1$&$\Delta L=2$&$\Delta L=3$&\\
\hline
$1010110100|0000$&$1010110010|0000$&$1010110001|0000$&$1010110000|1000$&\\
                 &$1010101100|0000$&$1010101010|0000$&$1010101001|0000$&\\
                   &               &$1001110010|0000$&$1001110001|0000$&\\
                                 & &$1010011100|0000$&$1010100110|0000$&\\
                   &               &$0110101100|0000$&$1010011010|0000$&\\
                                     & &             &$1001101010|0000$&\\
                                     & &             &$0110101010|0000$&\\
                                     & &             &$0110011100|0000$&\\
                                     & &             &$011010110101100|0000$&\\
                                     & &             &$100111001110010|0000$ &
\end{tabular}
\end{ruledtabular}
\end{table*}

\newpage
\section{III. Results for finite-layer thickness}

In a two dimensional system, the ideal Coulomb interaction between  electrons
has $V(q)=1/q$. The finite thickness in the normal direction  of an
experimental quantum Hall system modifies the short-distance part of the ideal 2D interaction, yielding an effective ``softer''
electron-electron interaction.
Here we include this non-zero thickness effect in the Coulomb interaction through the standard Fang-Howard model \cite{Yoshioka}.
The Fang-Howard model can faithfully describe two dimensional heterostructure in FQH experiments, which
assumed that the charge distribution normal to the x-y plane takes the form of variational wave function $\eta(z)=b^{-3/2} z \exp(-z/b)$,
where $b$ is the parameter giving the effective width of wavefunction in $z$-direction.
The effective electron-electron interaction is then written as \cite{Yoshioka}
\begin{eqnarray}
V(q)= \frac{1}{q} \frac{8+9qb+3q^2b^2}{8(1+qb)^3}.
\end{eqnarray}

\begin{figure}[!htb]
 \begin{minipage}{0.8\linewidth}
 \centering
 \includegraphics[width=4.0in]{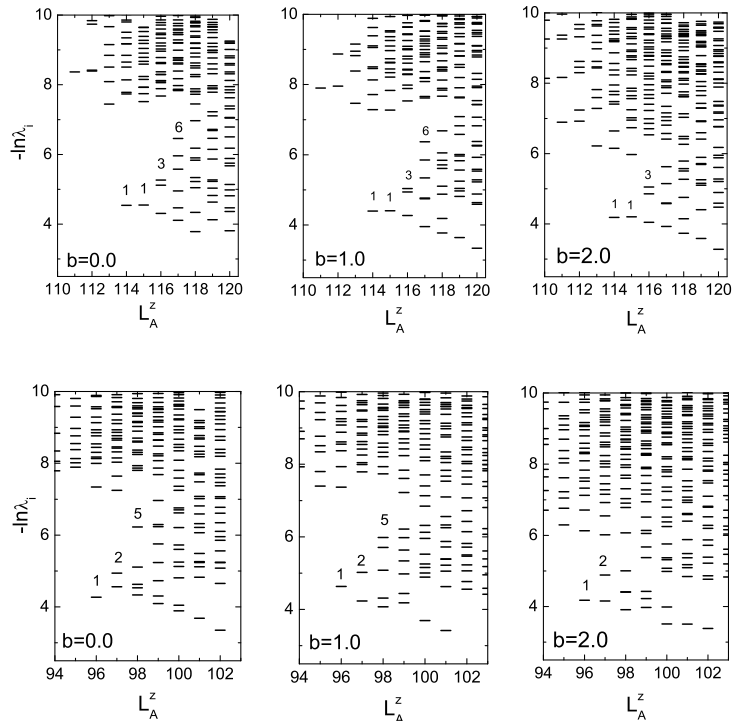}
 \end{minipage}
 \caption{Oribital ES of the groundstate for partition with $3n$ (top) electrons and $3n-1$ electrons (bottom), for different layer thickness parameter
 $b=0.0$, $b=1.0$ and $b=2.0$. The calculation is based on the pseudopotential obtained from plain geometry.
 The system size is $N_e=24$ and $N_s=37$.
 The first four degeneracy pattern are labeled by numbers.} \label{ES_plain}
\end{figure}

To study the effect of the finite-layer thickness, we use the pseudopotentials obtained from the infinite planar geometry (Eq. \ref{pseudo_disk})
where the finite-layer thickness effect can be  more conveniently obtained, although we study the quantum Hall systems  on the sphere geometry.
As noted before \cite{Peterson2008}, the pseudopotentials in the spherical geometry approach those in the planar geometry
if the spherical radius is taken to infinity in the thermodynamic limit.
To support the above statement, we present the entanglement spectrum (ES) at $b=0$, as shown in Fig. \ref{ES_plain}.
The countings in the first several momentum sectors match the prediction from RR $k=3$ state.
The similar picture obtained from the pseudopotential from planar geometry also indicates that
the ES is robust  and insensitive to the details of the pseudopotential form.
We further  show the result of nonzero layer thickness, in Fig. \ref{ES_plain}.
The counting for the first four momentum sectors ($\Delta L \leq 3$) are $1,1,3,6$ for partition with $3n$ electrons in the subsystem.
Examining the ES for different layer thickness,
it is found that the ES deviates from the expected counting in $\Delta L=3$ around $b\approx2.0$.

\begin{figure}[!htb]
 \includegraphics[width=0.65\linewidth]{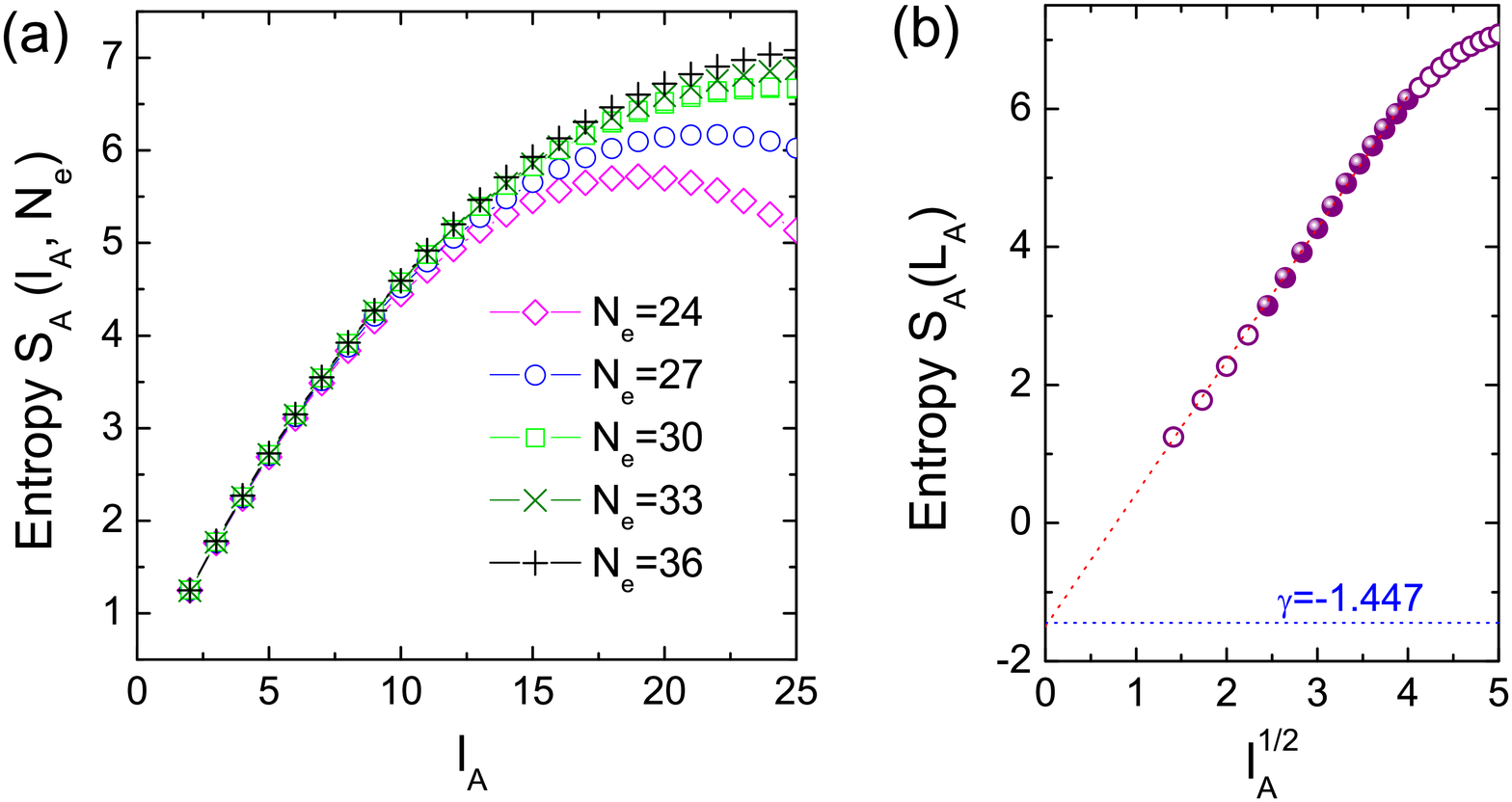}
 \caption{(a). Entanglement entropies with orbital partitioning $S(l_A,N_e)$,
for various system sizes $N_e$, where $l_A \le \frac{1}{2}N_e $
is the number of orbitals abive the cut
in the northern  hemisphere. (b). Scaling of entanglement entropies
to $S(l_A,N_e=36)=-\gamma+\alpha\sqrt{l_A}$ based on the solid circles
(O the sphere, The length of the cut is proportional to $\sqrt{l_A}$  \cite{Haque2006}).
The dashed blue line is the theoretical $\text{RR}_{k=3}$ TEE  value $\gamma\approx 1.447$.
The open circles were discarded in the extrapolation because they represent
very small subsystems ($l_A<5$) and violate the area law \cite{Haque2006,Estienne2014}
and the finite-size saturation effect ($l_A>16$).
} \label{entropy}
\end{figure}

\section{IV. Topological Entanglement Entropy}
For a two-dimensional gapped topologically-ordered state, the
entanglement entropy $S_A(l_A)$ for a subsystem A with a finite
boundary length $l_A$ is given by $S_A(l_A)=\alpha l_A - \gamma$,
where the TEE $\gamma$ is related to the total quantum dimension
$\mathcal{D}$ by $\gamma = \ln \mathcal{D}$ \cite{Kitaev2006,Levin2006}.
Since $\mathcal{D}$ contains the information about the quasiparticle
content, the  TEE
can determine whether a given topological phase belongs to
the universality class of a given topological field theory.

Fig. \ref{entropy}(a)
shows numerically-calculated orbital-cut entanglement entropy $S(l_A,N_e)$
as a function of the number of orbital ($l_A$) in the northern hemisphere for different system sizes ($N_e$).
The initially-increasing parts
of $S(l_A,N_e)$ reflect the physics of the macroscopic state, while
the downward curvature is a finite-size effect\cite{Haque2006,Zozulya2007,Estienne2014}.
With the help of the DMRG, we can obtain reliable entropies for $l_A \leq 16$,
because the entropy for a given $l_A\leq 16$ is
nearly saturated as $N_e$ increases from $30$ to $36$, as shown
in Fig. \ref{entropy}(a). In Fig. \ref{entropy}(b), we extract the  TEE (red line),
based on the raw data from $N_e=36$. The
obtained TEE is $\gamma=1.491\pm0.091$
(If we perform the extrapolation based on different system sizes, similar results are obtained:
for example, for $N_e=33$, we get $\gamma\approx 1.519 \pm 0.081$).
This  is consistent with
the theoretical value  $\ln \mathcal{D}=\ln \sqrt{5(1+\phi)}\approx1.447$ for RR $k=3$ state,
where each non-Abelian Fibonacci anyon quasiparticle sector
contributes quantum dimension $d_F=\phi=(\sqrt{5}+1)/2$ ($\phi$ denotes the Golden Ratio).
The appearance of $d_F=\phi$ is a signal of the emergence of Fibonacci anyon quasiparticles,
indicating two Fibonacci quasiparticles may fuse into
the  identity or into one Fibonacci quasiparticle.
This exotic property
makes Fibonacci quasiparticles   capable of universal quantum computation,
where all the quantum gates can be operated
and measured by braiding Fibonacci anyons \cite{Nayak}.

\section{V. Entanglement spectrum at $\nu=12/5$}
\begin{figure}[!htb]
 \begin{minipage}{0.8\linewidth}
 \centering
 \includegraphics[width=4.0in]{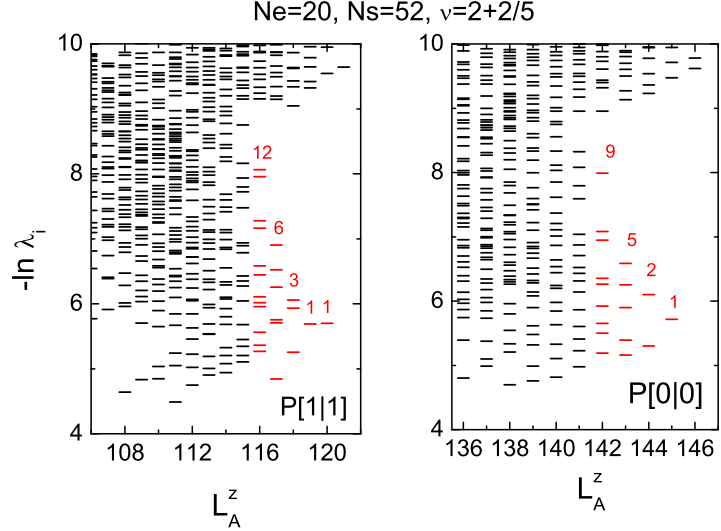}
 \end{minipage}
 \caption{Orbital ES of the GS for partition $P[1|1]$ and $P[0|0]$ for FQH $\nu=12/5$ state.
 The system size is $N_e=20$ and $N_s=52$.
 The first four multplicity-patterns are labeled by numbers.} \label{ES_PH}
\end{figure}

Here we show the entanglement spectrum for the groundstate at $\nu=2+2/5$, which is the
particle-hole (PH) conjugate state of $13/5$ state. In spherical geometry,
the highest density ``root configuration'' for the PH-conjugate RR $k=3$ state
has a pattern of ``0001100011000\ldots 11000''.
Consequently, there are also two distinct ways of partitioning, with
$2n-1$ (labeled as $P[1|1]$) and $2n$ ($P[0|0]$) electrons in the subsystem.
In Fig. \ref{ES_PH}, we show the ES of two partitions for $N_e=20$.
For $P[1|1]$ the leading ES displays the sequence of multiplicities  $1,1,3,6,12$ in the first five
momentum sectors $\Delta L=0,1,2,3,4$.
For $P[0|0]$, ES shows the multiplicity pattern  $1,2,5,9$ in the $\Delta L=0,1,2,3$ momentum sectors.
The ES picture found is exactly the same as that of the $\nu=13/5$ state, taking into account PH conjugation so that
$P[0|0]$ ($P[1|1]$) for $\nu=12/5$ relates to $P[1|1]$ ($P[0|0]$) for $\nu=13/5$.
The  low-lying ES structure is that predicted by the $SU(2)_3$ CFT for
the RR $k=3$ state,
and provides a fingerprint of the topological order for the groundstate at $\nu=12/5$.

%
%
%
%

\section{VI. Infinite DMRG on cylinder geometry}
The main results in this paper have been obtained with finite DMRG calculations
in   spherical geometry.
Spherical geometry simulations are  efficient for  calculating  the groundstate energy and related neutral gap.
The corresponding ES can  also be obtained. Nevertheless, one drawback of spherical geometry is that
one needs to select a ``shift'' value  $\mathcal{S}$ in the calculation. Usually, this value is determined by
some  empirical knowledge of the model wavefunction or one needs to compare results using different shifts.
Another disadvantage of spherical geometry is that the sphere has genus zero  so that it is
not suitable for discussing the topological degeneracy for topological ordered state.

An alternative strategy   is to  treat the cylinder geometry using
ithe infinite DMRG algorithm \cite{Zatel2013,Zatel2015,McCulloch}.
Here, we briefly introduce our implementation of infinite DMRG.
In the infinite-size DMRG, we first start from a small system size.
Then we insert several orbitals (for $\nu=13/5$, we add ten orbitals each time) in the center, and optimize the energy by sweeping over the inserted orbitals.
After the optimization, we absorb the new orbitals into the original
existing system (here $5$ orbitals are added to the left
system  and $5$ to the right environment)
and get the new boundary Hamiltonian.
We repeat these insertion, optimizing and absorption procedures until both  energy and entropy convergence are achieved by keeping a large number of states ($M$).
There are some  advantages of infinite DMRG over the finite DMRG.        
Compared with the finite DMRG simulation, infinite DMRG grows the system by several orbitals at each iteration and only sweeps the
inserted part, thus the computational cost is significantly reduced.
In the infinite DMRG algorithm, we do not need to set a ``shift'' $\mathcal{S}$ in the calculation.
One can also access  different topological sectors by randomizing the initial DMRG process.
Details of the realization of infinite DMRG on cylinder geometry and the related benchmark for model Hamiltonian 
will be given elsewhere. Here we present some details related to the results  shown in the main text.

Working on the cylinder geometry, we choose the Landau gauge $\vec{A}=(0,Bx) $, which conserves the y momentum around the cylinder.
The single electron orbitals in N-th Landau level are:
\begin{equation}
\psi_{N,j}(x,y)=\left(\frac{1}{2^N N! \pi^{1/2} L_y l}\right)^{1/2} exp[i\frac{X_j}{l^2}y-\frac{(X_j-x)^2}{2l^2}]H_{N}(\frac{X_j-x}{l})
\end{equation}
where $X_j=\frac{2\pi l^2}{L_y} j, j=1,2,\ldots,N_s$ is the center in x axis and $l$ is the magnetic length.
$H_N(x)$ is the Hermite polynomial.

Introducing the destruction (creation) operator $a_{N,j}(a^{\dagger}_{N,j})$ for $\psi_{N,j}$, the Coulomb interaction can be written as
\begin{equation}
H_C=\sum\limits_{N_1,\ldots,N_4} \sum\limits_{j_1,..,j_4} V_{N_1,j_1,\ldots,N_4,j_4} a^{\dagger}_{N_1,j_1}  a^{\dagger}_{N_2,j_2}  a_{N_3,j_3}  a_{N_4,j_4}
\end{equation}
where the Coulomb matrix elements are
\begin{equation}
V_{N_1,j_1,\ldots,N_4,j_4}=\frac{1}{2} \int d\textbf{r}_1 \int d\textbf{r}_2  \psi^{*}_{N_1,j_1}(\textbf{r}_1) \psi^{*}_{N_2,j_2}(\textbf{r}_2)  V(\textbf{r}_1,\textbf{r}_2) \psi_{N_3,j_3}(\textbf{r}_2) \psi_{N_4,j_4}(\textbf{r}_1)
\end{equation}
If we only consider the second Landau level (setting $N_i=1$), we can rewrite  the Hamiltonian  as
\begin{equation}
H_C= \sum_{l}\sum\limits_{n\geq0,m>0} V(m,n) a^{\dagger}_{l} a_{l+n}  a_{l+m+n} a^{\dagger}_{l+m+2n} + h.c.,
\end{equation}
where $V(m,n)$ is the matrix element derived from the modified Coulomb interaction.
In this work, when studying on cylinder geometry, we choose the form of the modified Coulomb interaction as \cite{Zatel2015}
\begin{equation}
  V(\textbf{r}_1,\textbf{r}_2) = \frac{1}{|\textbf{r}_1-\textbf{r}_2|} e^{-\frac{(\textbf{r}_1-\textbf{r}_2)^2}{\xi^2}}
\end{equation}
This is the form suitable for DMRG calculations  on the cylinder.
In the implementation, we kept  all Coulomb interaction terms $|V(m,n)|>10^{-6}$ within the truncated range $n<4\xi,m<\xi L_y/2$.
We have checked that the physical quantities remain qualitatively unchanged
when the truncation range is varied.
Here we would like to point out that one of  the important advantages
of the infinite DMRG is that it can  easily  deal with this type of the
Coulomb interaction in the cylinder geometry.
In traditional finite-size  DMRG in cylinder geometry,
an additional one-body potential $U(x)$ is needed to avoid the
electrons becoming trapped at the two ends of the finite cylinder,
when studying the systems with Coulomb interaction between electrons.
The infinite DMRG naturally overcomes this issue since it can access the actual results near the center by sweeping and edge effect should be suppressed when the length of the cylinder grows
long enough to reach the fixed point for the state  on the infinite cylinder.


To access the topologically different groundstates on cylinder using infinite DMRG,
we repeatedly start the infinite DMRG simulation for a given system size $L_y$ for several times.
In all calculations, we do not presume any empirical knowledge from model wavefunction.
To be more explicit, we do not set a seed-state or an orbital configuration according to the root configuration in the initial DMRG process.
We find that, in the relatively larger systems ($L_y\in[21,24]$), the system will automatically select
one of the two groundstates $|\Psi_1\rangle$ and $|\Psi_{\phi}\rangle$ with almost equal probability.
In the smaller system size ($L_y\in[18,20]$), the system has larger probability to fall into $|\Psi_{\phi}\rangle$ than $|\Psi_{1}\rangle$.
In all cases, we have double checked both of the groundstates are stable and robust with changing the parameters in DMRG calculation.
Once one groundstate has been developed, the groundstate is robust against increasing keep states or increasing the cylinder length in DMRG.
For example, in Fig. \ref{idmrg_test}, we show the entropy evolution of one groundstate in $|\Psi_{1}\rangle$ (red dots) and the other one in $|\Psi_{\phi}\rangle$ (black dots). The datas come from two independent infinite DMRG simulations.
At the point marked by $M=8000$, we change the keep states from $m=7000$ to $m=8000$ and keep $m=8000$ for all steps after.
The entropy of the two groundstates slightly increases with the increase of the keep states and the system length.
There is no sign of tunnelling between the two groundstates in our infinite DMRG calculations. Physically,
the tunnelling between two topological groundstates is forbidden since each topological groundstate hosts a
well-defined anyonic flux line and the changing the global anyonic flux is energetically expensive.
Furthermore, to check the two-fold groundstate degeneracy are complete, we try to start several infinite DMRG simulations
with different random initializations. It is found that all simulations will randomly fall into $|\Psi_{\phi}\rangle$ or $|\Psi_{1}\rangle$.
Thus we confirm the two-fold groundstates are indeed complete.
Within randomizing the initial processes, although the two groundstates can be distinguished by characteristic orbital entanglement spectrum (as shown in main text), we find a small energy fluctuation and entropy fluctuation for each groundstate when we  keep the same order of states in DMRG calculation.
The entropy fluctuation is much smaller than the entropy difference $\Delta S\approx \ln \phi \approx 0.47$ between two topological sectors.
We demonstrate the energy and entropy fluctuations as the error bars in Fig. 4(c-d) in the main text.
The relatively larger uncertainty of entropy in system size $L_y\leq20$ may result from:
finite-size effect or that the groundstate at Coulomb point is very close to the phase boundary to
a charge-density-wave (strip) state \cite{Rezayi2009}. To clarify the latter possibility, we measure the mean orbital
occupation number $\langle n_k \rangle$ in the middle part of infinite cylinder.
All of the groundstates have uniform occupation number with tiny fluctuation ($\Delta n_k <10^{-2}$).


\begin{figure}[!htb]
 \begin{minipage}{0.95\linewidth}
 \centering
 \includegraphics[width=2.5in]{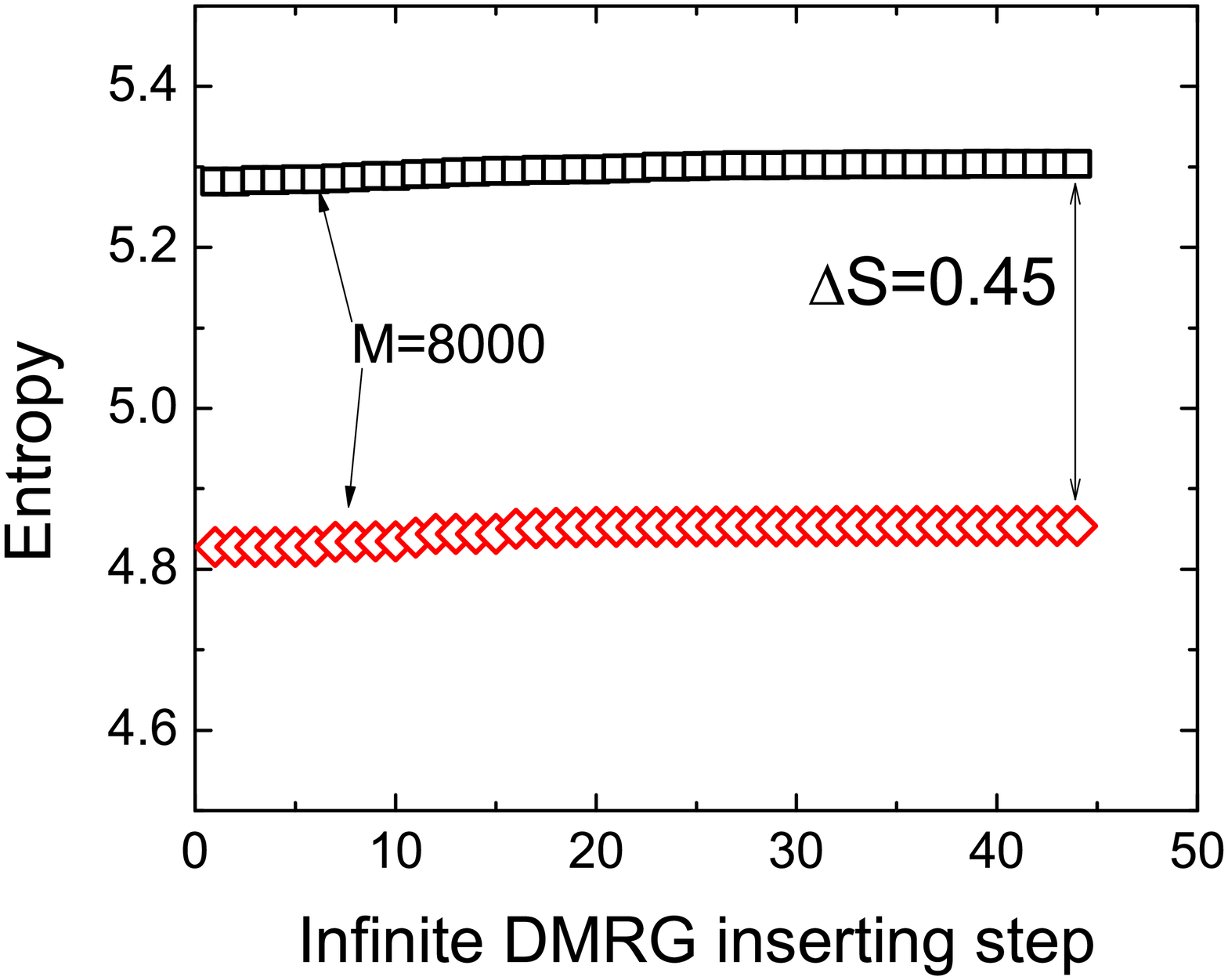}
 \includegraphics[width=3.5in]{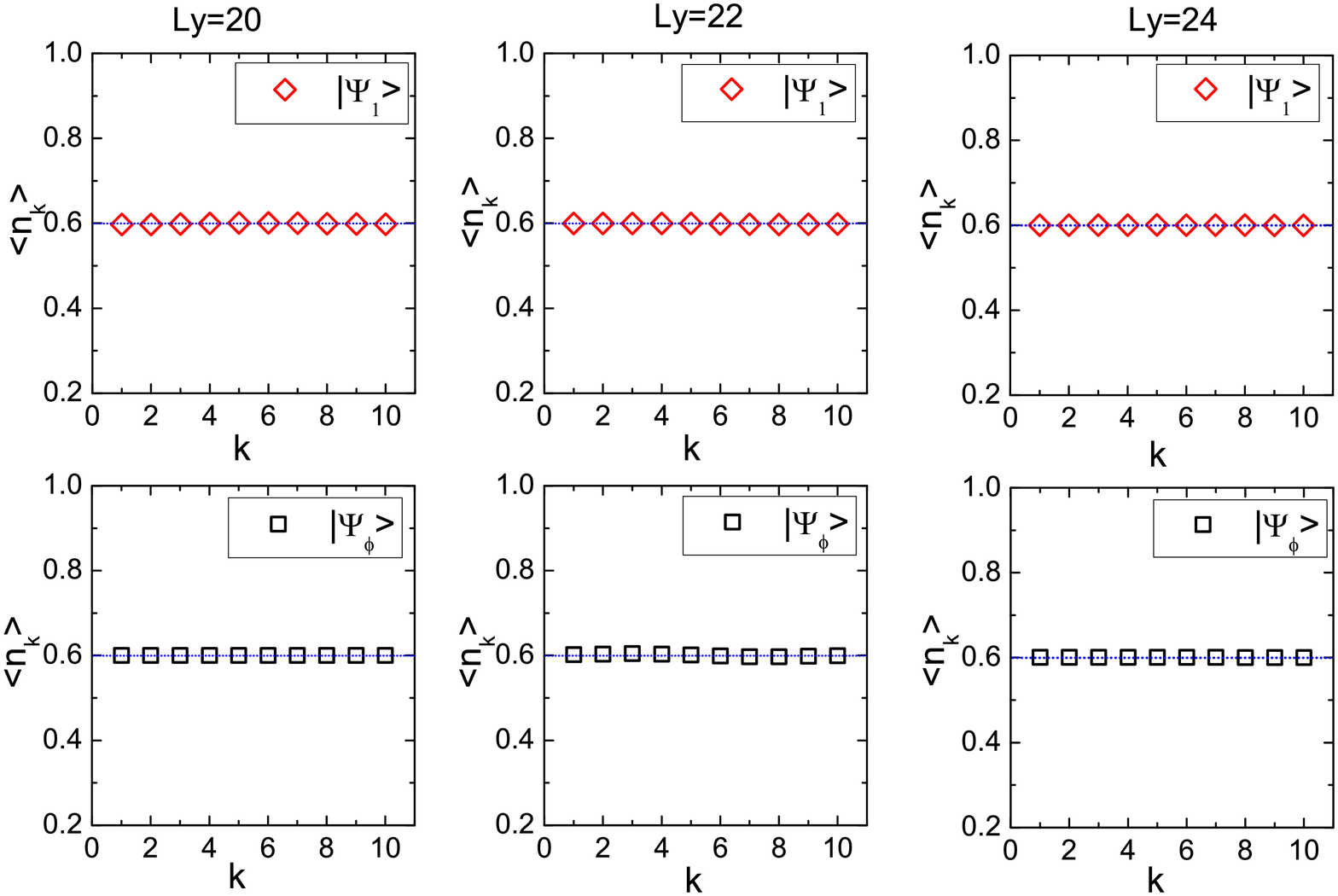}
 \end{minipage}
 \caption{(Left) The entropy evolution of one groundstate realizing $|\Psi_{\phi}\rangle$ (black dots) and $|\Psi_{1}\rangle$ (red dots).
  These two groundstates come from two independent infinite DMRG simulations.
  We omit the infinite DMRG steps before we reach a nearly converged groundstate with keeping $M=7000$.
  We change the keep state from $M=7000$ to $M=8000$ at the point as arrow marked.
  Each infinite DMRG step means inserting ten orbitals in the middle of the cylinder and sweeping until entropy converged.
  Here the system size is $L_y=24 l_B$.
  (Right) The mean orbital occupation number for two topological sectors on different system sizes. The dotted blue line shows $\nu=3/5$. } \label{idmrg_test}
\end{figure}

\end{appendices}
\end{widetext}


\begin{thebibliography}{99}
\bibitem{note1}
The charged gap is out of the scope of current study.
\bibitem{note3} The ES results shown in this paper give an unambiguous confirmation of RR$_3$ state as the ground state at $\nu=12/5$ and $13/5$. Based on this measurement, we can also exclude the possibility of the Jain CF state, because Jain state is expected to show $1,2,5,...$ degeneracy pattern in ES. Unfortunately, a direct exclusion of other candidates such as Bonderson-Slingerland (BS) state or bipartite CF state is unavailable, since it is still unknown the edge theory of these proposals. A further study of their excitations or edge theories will be necessary to distinguish between these proposals.
\end{thebibliography}
\end{document}